# Observing the Brownian motion in vibro-fluidized granular matter

G. D'Anna[1], P. Mayor[1], A. Barrat[2], V. Loreto[3] and F. Nori[4,5]

[1] *Institut de Physique de la Matière Complexe, Faculté des Sciences de Base, Ecole Polytechnique Fédérale de Lausanne, CH-1015 Lausanne, Switzerland*

[2] *Laboratoire de Physique Théorique, Unité Mixte de Recherche UMR 8627, Bâtiment 210, Université de Paris-Sud, 91405 Orsay Cedex, France*

[3] *Università degli Studi di Roma La Sapienza, Dipartimento di Fisica, and INFM, Center for Statistical Mechanics and Complexity, P.le A. Moro 5, 00185 Rome, Italy*

[4] *Frontier Research System, The Institute of Physical and Chemical Research (RIKEN), Wako-shi, Saitama 351-0198, Japan*

[5] *Center for Theoretical Physics, Department of Physics, CSCS, University of Michigan, Ann Arbor, Michigan 48109-1120, USA*

**At the beginning of last century, Gerlach and Lehrer[1,2] observed the rotational Brownian motion of a very fine wire immersed in an equilibrium environment, a gas. This simple experiment eventually permitted the full development of one of the most important ideas of equilibrium statistical mechanics: the very complicated many-particle problem of a large number of molecules colliding with the wire, can be represented by two macroscopic parameters only, namely viscosity and the temperature. Can this idea, mathematically developed in the so-called Langevin model and the fluctuation-dissipation theorem[3,4], be used to describe systems that are far from equilibrium? Here we address the question and reproduce the Gerlach and Lehrer experiment in an archetype non-equilibrium system, by immersing a sensitive torsion oscillator in a granular system[5,6] of millimetre-size grains, fluidized by strong external vibrations. The vibro-fluidized granular medium is a driven environment, with continuous injection and dissipation of energy, and the immersed oscillator can be seen as analogous to an elastically bound Brownian particle. We show, by measuring the noise and the susceptibility, that the experiment can be treated, in first approximation, with the same formalism as in the equilibrium case, giving experimental access to a "granular viscosity" and an "effective temperature", however anisotropic and inhomogeneous, and yielding the surprising result that the vibro-fluidized granular matter behaves as a "thermal" bath satisfying a fluctuation-dissipation relation.**

Before describing the experiment in the granular medium we briefly recall the Langevin formalism for the Brownian motion of a torsion oscillator immersed in a usual liquid at temperature $T$. This is a canonical problem of statistical mechanics[4]. The Langevin equation of the oscillator is $I(d^2\theta/dt^2) + \alpha(d\theta/dt) + G\theta = C_{ext}(t) + R(t)$, where $\theta$ is the angular deflection, $I$ is the moment of inertia of the oscillator, $\alpha$ the friction coefficient determining the viscous torque, $G$ the fibre elastic torsion constant, $C_{ext}(t)$ the external torque, and $R(t)$ a randomly fluctuating torque, assumed to be a Gaussian white noise of zero mean. This equation is valid on timescales large compared to the correlation time of the random torque.

The Langevin equation can be solved by Fourier transformation. Without the external torque ($C_{ext}(t) = 0$), using the Wiener-Khintchine theorem for stationary processes[4], one obtains the noise Power Spectral Density (PSD), i.e. twice the Fourier transformation of the auto-correlation function $<\theta(t)\theta(t')>$, which gives $S(\omega) = 2q/[I^2(\omega_o^2 - \omega^2)^2 + \alpha^2\omega^2]$ where $\omega_o = \sqrt{G/I}$ is the natural pulsation of the oscillator and $2q$ is the PSD of the random torque $R(t)$. Moreover, one has the relation $q = 2\alpha k_B T$, which guarantees that the system is in thermodynamic equilibrium and that the equipartition of energy holds. On the other hand, solving the Langevin equation with the external torque $C_{ext}(t)$, and focusing on timescales large compared to the correlation time of the random torque, yields the complex susceptibility $\chi(\omega) = \theta(\omega)/C_{ext}(\omega) = \chi'(\omega) - i\chi''(\omega)$ with $\chi(\omega) = 1/[I(\omega_o^2 - \omega^2) + i\alpha\omega]$. One finds that the ratio $S(\omega)\omega/(4\chi''(\omega)) = k_B T$ is proportional to the temperature of the liquid, and is independent of the oscillator characteristics, such as its mass or shape. This represents a formulation of the fluctuation-dissipation theorem[4] (FDT), expressed in the frequency space. Summing up, measuring the susceptibility and the noise PSD of the Brownian motion of the oscillator allows one to consistently determine the viscosity and the temperature of the liquid seen as an ideal thermal bath at equilibrium.

The natural question that arises at this point is how much of this formalism survives if the experiment is performed in a strongly non-equilibrium system, notably in a vibro-fluidized granular medium[5,6]. A granular system is an assembly of particles, such as sand grains or glass beads, interacting by contact forces and featuring a very large number of macroscopic degrees of freedom, corresponding in first approximation to the positions and velocities of all grains. However, such a system is not in "equilibrium" in the thermodynamic sense, since the thermal energy $k_B T$ at room temperature is too small to induce any macroscopic grain fluctuations. Nevertheless, due to the very large number of degrees of freedom, one expects that an analogy to the thermally-induced Brownian motion is possible when the system is externally driven and grain motion occurs by continuous injection of energy.

We have realized the experiment by immersing a torsion oscillator in a granular medium composed of glass beads (Fig. 1). The container, filled with the glass beads, is continuously shaken by a vertical vibrator, with a high-frequency filtered white noise, cut off below 300 Hz and above 900 Hz. Notice the important point that a filtered white noise is used in order to obtain a homogenous agitation of the granular medium, discarding undesired effects such as pattern formation, rolls, and other instabilities in the granular motion[5]. The purpose is not to provide *ab initio* a random torque with white noise spectrum to the oscillator. Indeed, we observe the motion of the oscillator in a low-frequency range (10 Hz to 50 Hz) compared to the applied vibration high-frequency range. The probe of the torsion oscillator, with various moments of inertia and shapes, is immersed at a depth $L$ from the granular surface. The oscillator is otherwise isolated from the container and the vibrator, and does not move in the vertical or horizontal directions. An accelerometer on the container is used to measure the acceleration spectrum, normalized to the acceleration of gravity, $A(\omega)$. We quantify the "intensity" of the external shaking by $\Gamma$, defined as the square root of the band power of $A(\omega)$ in the range about 1 Hz to 10 kHz, i.e. $\Gamma^2 = \int A(f)df$ integrated in the above frequency range, with $\omega = 2\pi f$. In the case of a sinusoidal vibration of amplitude $a_s$ and frequency $f_s = \omega_s/2\pi$, the normalized acceleration spectrum is $A(f) = \left[a_s(2\pi f_s)^2/g\right]^2 \delta(f - f_s)$ and one recovers the usual definition of the vibration intensity, i.e. $\Gamma = a_s\omega_s^2/g$. For sinusoidal vibrations, $\Gamma = 1$ is the lift-on threshold above which a single grain starts to "fly". In this work we employ vibration intensities up to $\Gamma \approx 15$.

We now use the oscillator to measure 1) the noise PSD, and 2) the susceptibility of the vibro-fluidized granular medium:

1) In absence of the external torque ($C_{ext}(t) = 0$) the immersed oscillator performs an irregular free angular motion, induced by the continuous interactions of the grains against the probe. The angular deflections $\theta$ are detected optically using a mirror fixed on the oscillator and, from the time-series of the angular deflections, $\theta(t)$ we obtain the noise PSD, $S(\omega)$, as shown in Fig. 2a for several values of $\Gamma$.

2) While the granular medium is vibrated at a given intensity $\Gamma$, a sinusoidal torque $C_{ext}(t) = C_e \sin(\omega t)$ is applied to the oscillator using a permanent magnet fixed on the oscillator and two external coils. The complex susceptibility at the given $\omega$ is obtained from $\chi(\omega) = \theta(\omega)/C_{ext}(\omega)$, where $\theta(\omega)$ and $C_{ext}(\omega)$ are the Fourier transforms of $\theta(t)$ and $C_{ext}(t)$, respectively. The amplitude $C_e$ of the external torque is small enough to be in the regime of linear response, i.e., we measure the linear susceptibility. The complex susceptibility $\chi(\omega)$ versus $\omega$ is obtained by sweeping the frequency of the sinusoidal torque. The modulus of the complex susceptibility, $|\chi(\omega)|$, for different intensities of vibration $\Gamma$, is shown in Fig. 2b.

With both the susceptibility and the noise PSD data, we are now in the position to check whether the analogy to the thermal-induced Brownian motion makes sense. First of all, the modulus of the complex susceptibility, $|\chi(\omega)|$, is fitted remarkably well by the susceptibility expression for the damped oscillator, $|\chi(\omega)| = \left[I^2\left(\omega_o^2 - \omega^2\right)^2 + \alpha^2\omega^2\right]^{-1/2}$, as shown in Fig. 2b. This means that the deterministic equation of motion, i.e. $I(d^2\theta/dt^2) + \alpha(d\theta/dt) + G\theta = C_{ext}(t)$, describes well the response of the immersed oscillator, and allows us to define a granular friction coefficient $\alpha$, or a granular viscosity $\mu \propto \alpha$. From the fitting parameters at different $\Gamma$ we also deduce that $\alpha \propto 1/\Gamma$, as shown in the inset of Fig. 2b.

Secondly, we can obtain the fluctuation-dissipation (FD) ratio $S(\omega)\omega/4\chi''(\omega)$ from our data. Fig. 3a shows the FD-ratio versus $f = \omega/2\pi$ at different $\Gamma$. The FD-ratio is surprisingly "flat", i.e. approximately independent of the frequency in the observed low-frequency range (especially compared to what has been measured in other non-equilibrium thermal systems, such as in laponite[7] and glycerol[8].) This means that the high-frequency driven agitation of the granular medium acts on the oscillator as a source of random torque with white spectrum, at least in the 10 Hz to 50 Hz range under investigation. Energy is thus injected at high-frequency, and spreads into a low-frequency white spectrum. A roughly flat FD-ratio also provides support for the existence of a fluctuation-dissipation relation in off-equilibrium driven granular steady states. The FD-ratio level can thus be used to define an effective temperature, $T_{eff}$. Fig. 3b shows the averaged FD-ratio levels, i.e. $k_B T_{eff}$, versus $\Gamma$. Fitting to a power-law we obtain $k_B T_{eff} \propto \Gamma^p$ with p close to 2, thus giving $T_{eff} \propto \Gamma^2$.

Obviously, this is only a first-order approximation. Close inspection of Fig. 3a apparently shows a small frequency dependence. However, an useful torsion signal is only detected around the natural frequency of the oscillator, limiting our accessible frequency range. In future work, the study of the frequency dependence of the FD-ratio over a large frequency range could give insights about possible energy "cascade" in vibro-fluidized granular matter.

In the frame of this first-order approximation, our results suggest a simply picture. Due to the complex dissipation processes between the grains, only a fixed fraction of the energy injected by the vibrator is effectively available as granular kinetic energy and is "sensed" by the oscillator. In fact, we notice that the order of magnitude of the thermal energy, i.e. $k_B T_{eff}$, as measured here, is consistent with realistic values of the mean kinetic energy per particle, as measured by grain-tracking methods, e.g., in ref. 9 and 10. Thus, the effective temperature $T_{eff}$ we measure seems to be related to the granular temperature, as usually defined in granular gases[5,6]. Notice that the measured granular friction coefficient $\alpha$ decreases by increasing the effective temperature, since $\alpha \propto 1/\Gamma$ and $T_{eff} \propto \Gamma^2$. This is the behaviour of liquids rather than gases. However, the oscillator sees an increasing effective temperature *and* a decreasing granular density as $\Gamma$ is increased. A similar effect has been reported by Zik et al. (ref. 11) by observing the mobility of a sphere immersed in a vibro-fluidized granular medium.

We now study the dependence of $T_{eff}$ on the oscillator properties, as well as on the characteristics of the granular medium, such as immersion depth and granular anisotropy. Fig. 3b addresses all these issues. The main panel of Fig. 3b shows data obtained using a conical probe with different moments of inertia. The upper inset of Fig. 3b shows data obtained using cylindrical probes with different sections and diameters. In all these circumstances it turns out that $T_{eff}$ is insensitive to the change of these parameters. This is an indication that $T_{eff}$ could be an intrinsic property of the granular medium.

The bottom inset of Fig. 3b shows data measured at different immersion depths, using conical and cylindrical probes. Two important aspects needs to be underlined: First, in a layer close to the surface the measured FD-ratio shows a strong frequency dependence, and it becomes "flat" (as in Fig. 3a) only at a sufficient immersion depth. (The FD-ratio becomes "flat" about at the depth where a minimum appears in the bottom inset of Fig 3b). Underneath the surface layer, we observe that $T_{eff}$ increases with the depth. This is possibly related to a slight decrease of the granular density with the depth, as observed in simulations[12] and measurements[13]. Regarding the behaviour of the surface layer itself, even though we cannot exclude a relation with density induced features (as observed in refs. 10,12 and 13), it ought to be explored more carefully. A second point is the difference between $T_{eff}$ measured by conical and cylindrical probes. This is a peculiar granular effect and it is probably related to the intrinsic anisotropy of the system under gravity. Although the diameter or the form of the horizontal section of the probe play no role, the angle of inclination of the surfaces exposed to the granular agitation, relative to the vertical axis, seems to be a relevant parameter. As reported in ref. 10, the granular temperature is anisotropic, corresponding to horizontal and vertical impulse components. One thus expects in particular the transfer of the vertical impulse into torsion momentum to depends on the geometry and the surface state of the probe, and this effect will be the object of further investigations. Nevertheless, for a given kind of probe, e.g. conical or cylindrical, the present experiment shows that one can satisfy the requirements for the Langevin approach to be applicable.

The most important finding obtained here is that it is possible to separate slow and fast degrees of freedom in vibro-fluidized granular matter and to model it, in the very spirit of the Langevin approach, in terms of two intrinsic parameters: an effective friction coefficient and an effective temperature. In this perspective the oscillator behaves indeed as a Brownian oscillator, and the vibro-fluidized granular matter can be seen as a macroscopic "thermal bath". We caution that the analogy is only formal, since the granular medium is never in equilibrium in the thermodynamic sense. The experiment also shows that the granular matter's mechanical anisotropy and inhomogeneity due to the gravitation play an important role, inducing complications that are not present in usual gas or liquid media, which can be assumed isotropic and homogenous.

Acknowledgements
We thank Zoltan Racz and Alessandro Vespignani for very useful discussions and comments.

LEGEND

**Figure 1** Sketch of the torsion oscillator immersed in a "viscous liquid". Here the "viscous liquid" is a granular medium fluidized by strong external vibration (see text for details). The granular material consists of glass beads of diameter 1.1±0.05 mm contained in a metallic bucket of 60 mm height and 94 mm diameter, filled to a height of 31 mm. The oscillator has a natural frequency of about 12 Hz, depending on the probe used. The main probe used is a cone of apex angle 120°, covered with a single layer of glass beads, glued on by an epoxy. Probes with different shapes have also been used, in particular cylindrical probes with various section and diameters, see Fig. 3. The probe is immersed at a depths $L$. The oscillator allows measurement of both the noise PSD and the complex susceptibility. An accelerometer permits to quantify the intensity of the external vibration, $\Gamma$. Notice that the oscillator does not move in the vertical or horizontal directions. For this reason, below $\Gamma \approx 1$ the oscillator acts as a fixed-displacement deformation point, and the study of the granular dynamics at very low $\Gamma$ requires a different approach (ref. 14).

**Figure 2** Noise and susceptibility for different vibration intensities. The probe is conical, immersed at about 11 mm from the surface. **a,** Noise PSD, $S(\omega)$, versus the frequency $f$, for different vibration intensities $\Gamma$. The intensity values are, from top to bottom, 11.6, 10.0, 7.3, 5.4, 3.7, 2.2, 1.5 and 1.0. **b,** Modulus of the complex susceptibility $|\chi(\omega)|$ versus the frequency $f = \omega/2\pi$, for different vibration intensities $\Gamma$. The intensity values are as above. Each curve (continuous line) is fitted (dashed line) with the expression $|\chi(\omega)| = \left[ I^2(\omega_o^2 - \omega^2)^2 + \alpha^2 \omega^2 \right]^{-1/2}$, with two free parameters, $\omega_o$ and $\alpha$. The moment of inertia of the oscillator is $I = 1.5 \cdot 10^{-6}$ kg m$^2$, and the applied torque amplitude is $C_{ext}(t) = 3.2 \times 10^{-5}$ N m. The parameter $\omega_o$ shows a small dependence on $\Gamma$: it features a small increase from 12 Hz to 18 Hz by reducing $\Gamma$. The Inset shows the friction coefficient $\alpha$, obtained from the fit to the curves $|\chi(\omega)|$ in the main panel, versus $\Gamma$. The dashed line is a power law $\alpha \propto 1/\Gamma$. The $1/\Gamma$ dependence is also systematically observed using susceptibility data obtained with probes of different shapes.

**Figure 3** FD-ratio and the effective temperature. **a,** The FD-ratio $S(\omega)\omega/(4\chi''(\omega))$, versus $f = \omega/2\pi$ for different vibration intensities $\Gamma$, obtained from Fig. 2. The intensity values are, from top to bottom, 11.6, 10.0, 7.3, 5.4, 3.7, 2.2, 1.5 and 1.0. **b,** The FD-ratio level (averaged from 10 Hz to 50 Hz), i.e., the effective temperature $T_{eff}$, according to the expression $S(\omega)\omega/(4\chi''(\omega)) = k_B T_{eff}$, versus $\Gamma$, from data in Fig. 3a (black open symbols), and from similar data obtained using a conical probe with triple moment of inertia (red symbols) and the same $L$. A power law fitted to the data gives $T_{eff} \propto \Gamma^p$, with p=2.1. The dashed line has equation $k_B T_{eff} = 3.5 \times 10^{-10} \Gamma^2$. The same FD-ratio levels are measured feeding the vibrator by a filtered white noise with a different central frequency (800 Hz, not shown). The FD-ratio level depends also on the tribological properties of the granular material (especially rich for glass, ref. 15). We have obtained reproducible data using aged (i.e., exposed long time to atmosphere) glass beads. Upper inset, the effective temperature $T_{eff}$ versus $I$, obtained using probes of different shapes: a cylinder of circular section of 5 mm diameter and polished surface, and a cylinder with "toothed wheel" section, with polished surfaces, as sketched. The immersion depth is about 21 mm. The same FD-ratio levels are also observed using circular section of 7 mm diameter. Lower inset, the effective temperature $T_{eff}$ at $\Gamma = 9$, versus the immersion depth of the oscillator, $L$, for conical (red symbols) and cylindrical (black symbols) probes (see text for details) (For clarity, only one symbol every four data values is marked). Notice that the precise position of the surface in unknown for a vibrated granular medium, but it is approximately at $L = 0$ (vertical dashed line). The bottom of the container is at $L = 31$mm. We caution that in the layer close to the surface (about 0<L<5 mm), the FD-ratio is anymore "flat", and we do not report data in this region.

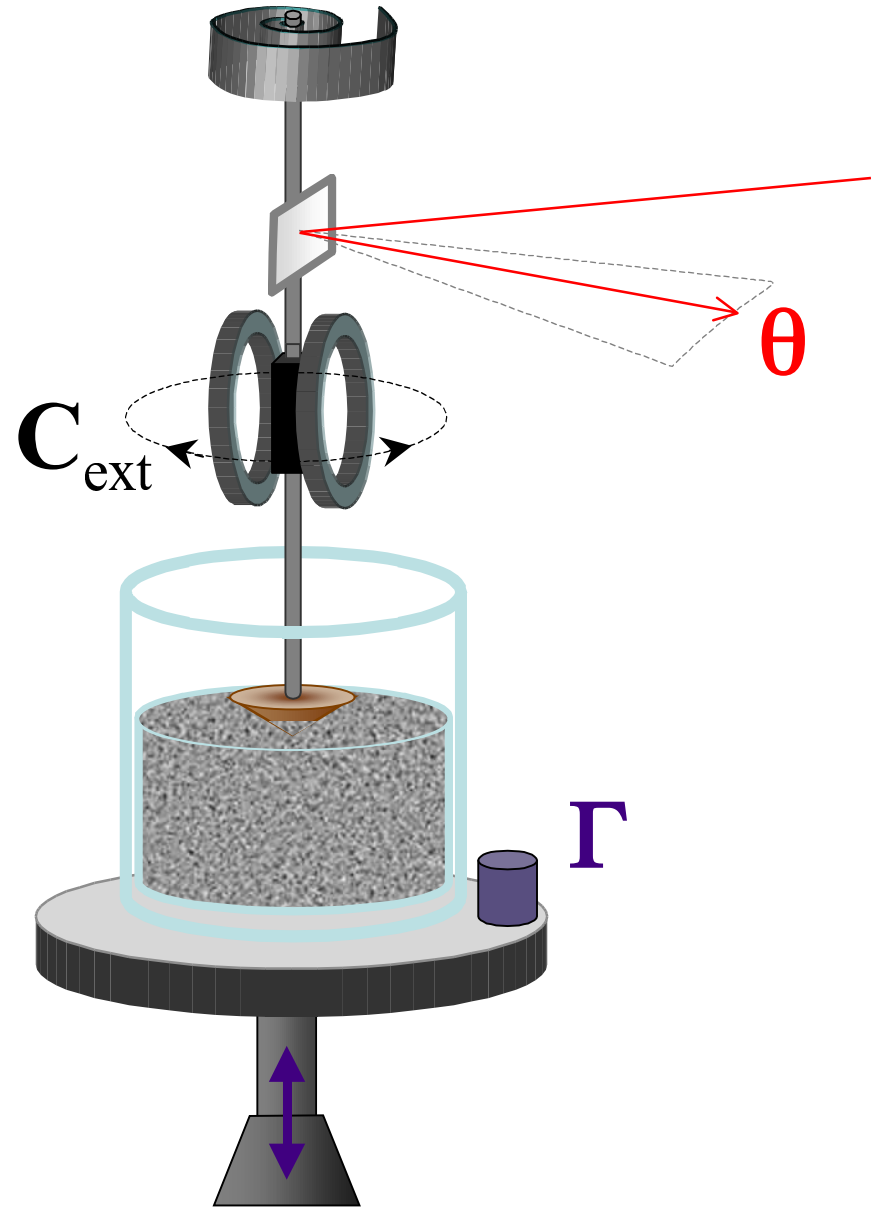

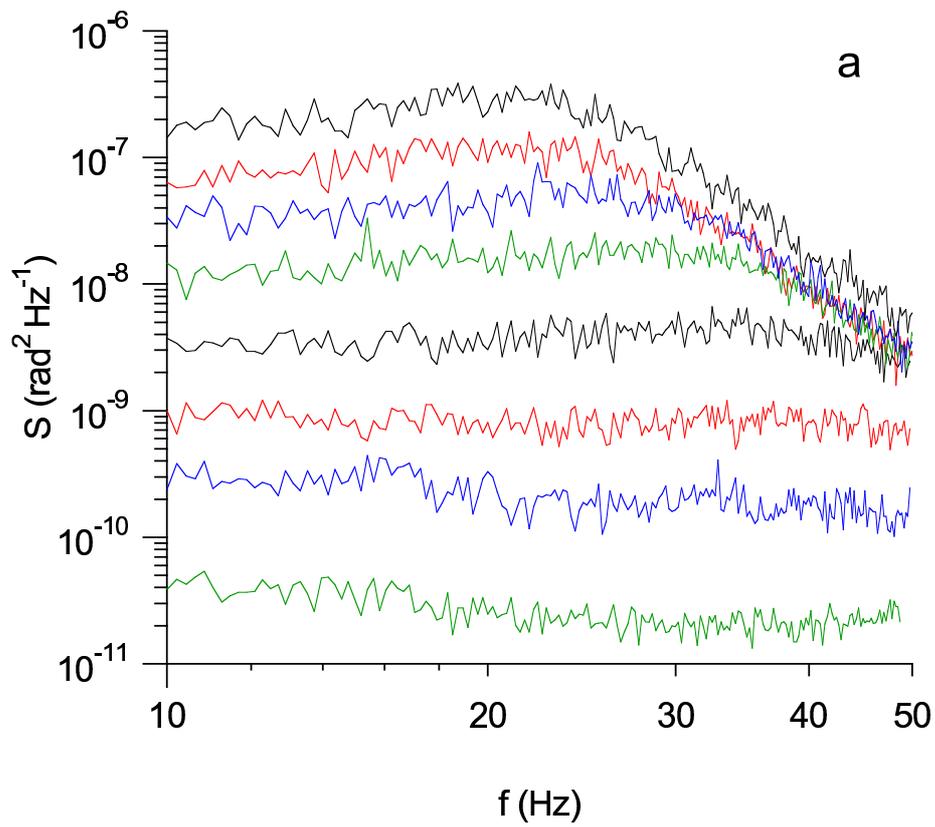
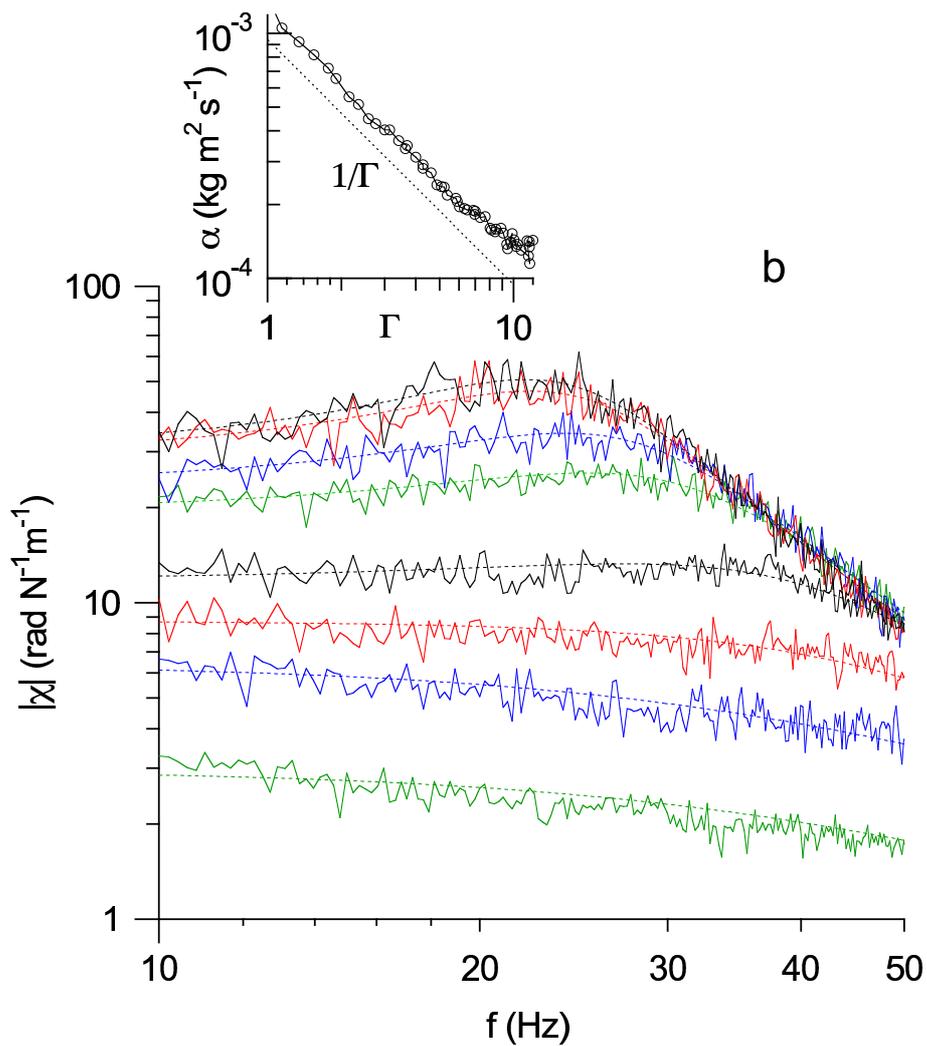

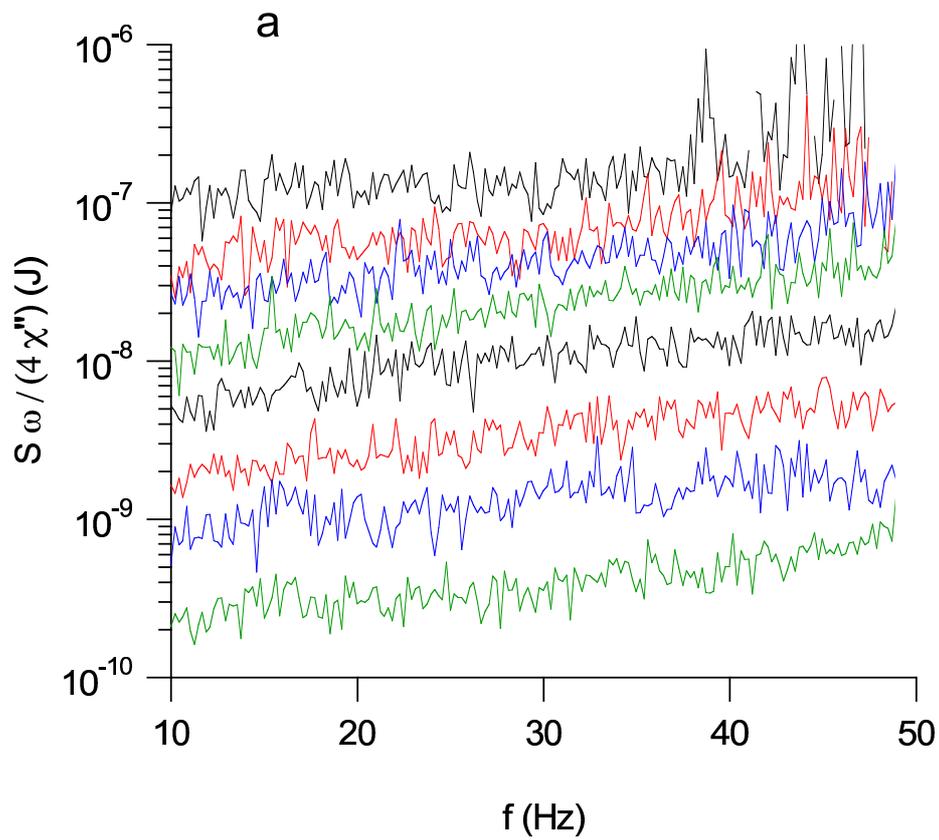

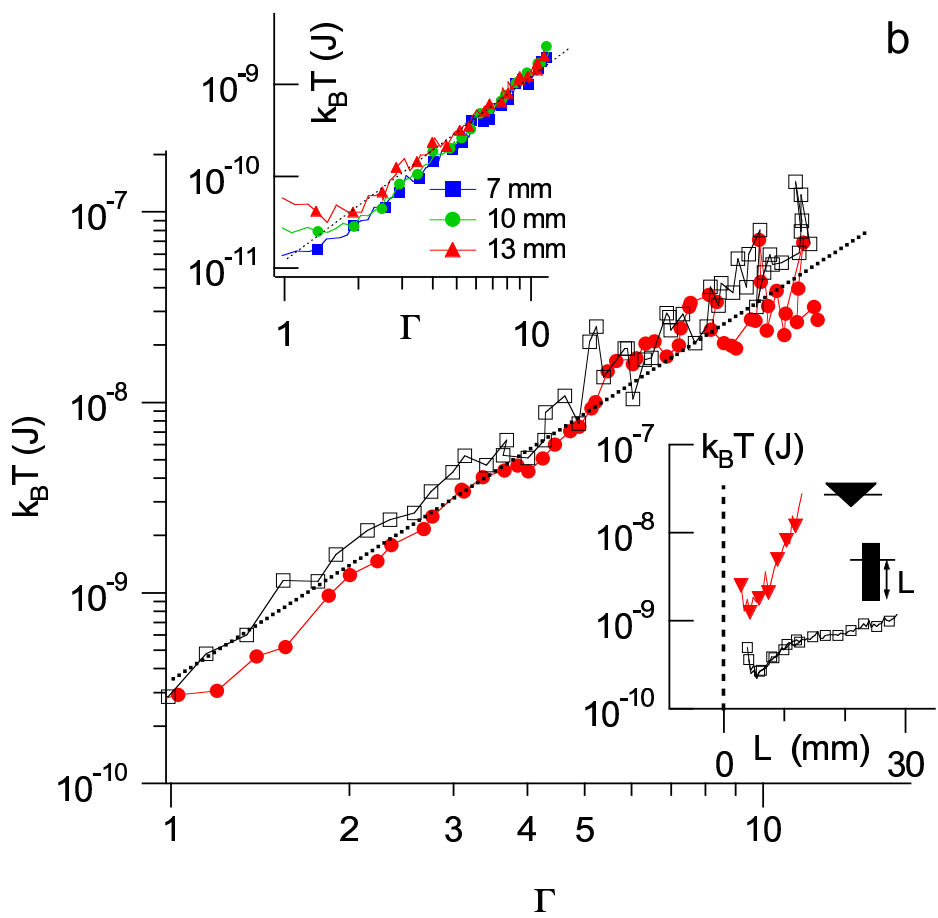